\documentstyle[11pt,epsf]{article}

\title{{\bf What Does the Balance Function Measure?}}

\author{{\bf Thomas A. Trainor}\\ \\
{\em CENPA 354290}\\
{\em University of Washington}\\
{\em Seattle, WA 98195}\\
{\em trainor@hausdorf.npl.washington.edu}}

\def\be{\begin{equation}}
\def\ee{\end{equation}}
\def\bea{\begin{eqnarray}}
\def\eea{\end{eqnarray}}

\newcommand{\xfig}[1]{\begin{center}
\mbox{\epsffile{#1}}
\end{center}}

\begin{document}
\maketitle

\begin{abstract}
The balance function has been proposed to `clock' hadronization by measuring the charge-dependent correlation length on rapidity. According to the proposal a narrowed balance function would imply reduced hadron diffusion time and therefore delayed hadronization due to a long-lived prehadronic state, a quark-gluon plasma or QGP. I examine this chain of argument in the context of hadronization and rescattering. I then consider the algebraic structure of the balance function and its systematic dependence on the charge correlation length on rapidity. I conclude that the width of the balance function cannot determine a time interval from hadronization to kinetic decoupling, nor can the width determine the existence or temporal extent of a QGP.
\end{abstract}

\section{Introduction}

Balance function measurements were motivated by predictions \cite{bass,jeon} of length-scale systematics for charge-dependent rapidity correlations in heavy ion collisions, according to which changes in correlation length are caused by thermal and diffusion effects marking the elapsed time from hadronization. Measurement of a charge-dependent correlation length on rapidity with the balance function would thus reveal the timing of hadronization, and detection of a reduced width of the balance function would indicate a long-lived prehadronic or QGP state \cite{bass}. It was suggested to vary collision centrality and energy to look for such narrowing, a sudden drop in the balance-function width signaling QGP formation \cite{jeon}.

According to \cite{jeon} both the balance function and net-charge fluctuations are sensitive to charge-dependent longitudinal momentum correlations. Production of a QGP of significant space-time extent would imply substantial production of new charge late in the collision, reducing the time hadron charge pairs have to diffuse apart by rescattering. A reduced balance function width or `tight correlation' would thus provide evidence of a QGP. By a related argument the presence of a QGP would be signaled by unusually small net-charge fluctuations \cite{qgp-charge}. The theoretical hypothesis and balance function width interpretation depend on two aspects: 1) the relation of a long-lived QGP to the charge-dependent correlation length on rapidity and 2) the relation of the balance function width to that correlation length. The interpretation of the $rms$ width of the balance function relative to the charge-dependent correlation length and the structure of the theoretical motivating argument are the subjects of this paper.

I first examine the physical arguments motivating the balance function measurement, then review some analyses of elementary collisions which served as prototypes for the balance function, summarizing the methods and the understanding of hadron charge correlations which resulted. I restate the balance function definition in a more detailed algebraic form and thereby relate it to the autocorrelation distribution. I adopt a simple model of the charge-dependent autocorrelation as observed in heavy ion collisions and use this to investigate balance function width systematics relative to the charge-dependent correlation amplitude and length. 

\section{Physics of the Balance Function}

According to the theoretical motivation of the balance function in \cite{bass,jeon}, the appearance of quark pairs as hadrons signals the end of the QGP phase and the start of hadron diffusion (rescattering). To say that hadronization is `late' implies that a QGP existed for an extended prior time interval. Diffusion alters hadron charge correlations: separation of correlated charge pairs in space increases monotonically with diffusion time. Schematically, the argument in \cite{bass} is: 1) Hadron charge pairs are created at single points in space-time; 2) thermal motion, diffusion and axial Hubble expansion map these points to a peaked distribution on relative rapidity, the peak width (correlation length) increasing with diffusion time; 3) the balance-function width measures this correlation length; thus, 4) if the balance-function width is reduced the time from hadronization to decoupling is less and the time before hadronization is greater, indicating an extended QGP lifetime. 

\subsection{Details of theoretical argument}

The motivating language for the balance function implies a fixed time interval between first nuclear contact and kinetic decoupling, divided into two parts by hadronization. The existence and extent of the prior QGP interval are the nominal objects of the balance function analysis. During hadronization quark and/or hadron charge pairs are said to appear at nearly the same instant, and hadron charge pairs are created at zero space-time separation \cite{bass}, which then defines the initial space-time autocorrelation of hadron charge pairs in this picture. 

The autocorrelation on space difference variable $z_\Delta$ maps to the autocorrelation on rapidity difference $y_\Delta$ $via$ the phase-space correlation of axial Hubble flow or Bjorken expansion. The width of the corresponding structure on $y_\Delta$ is determined by several factors, primarily thermal velocities at kinetic decoupling and diffusion (rescattering) time from hadronization to kinetic decoupling. Cooling by isentropic expansion and diffusion time are both indicated by changes in the charge-dependent {\em correlation length} on rapidity according to the argument.

The balance function, based on observables formulated in \cite{epem}, was designed to quantify these rapidity correlations.  It is said to describe the conditional probability that a charge in one rapidity bin is accompanied by a charge of opposite sign in another bin. If the balance function width decreases (due to reduced diffusion time) the time interval prior to hadronization must have increased. `Late' or `delayed' hadronization inferred from the balance-function width thus implies a long-lived QGP. The authors of \cite{bass} conclude ``...it seems clear that the canonical picture of a heavy ion reaction, quark-gluon plasma formation followed by late-stage hadronization, should have a clear signature in the balance functions.''

\subsection{Critique of theoretical argument}

\begin{itemize}

\item Fixed total interval: The theoretical argument relies on implicit {\em duration complementarity} between the hadronic time interval and the prehadronic or QGP interval. Reduction of the former implies increase of the latter. The fixed total interval is conveyed by language like `late-stage' or `delayed' hadronization, which suggests an extended QGP lifetime. The implication is not justified.

\item Point-like pair creation: The assumed point-like space-time autocorrelation for oppositely-charged hadrons at formation is inconsistent with the fragmentation picture contained in the Lund model \cite{lund}. The actual hadron autocorrelation at formation should have a non-zero width representing QCD string physics (in the 1D case), which hadronic diffusion, thermal velocities and other contributions then modify in the projection to rapidity difference. Rapidity correlation lengths in elementary collisions, best reflecting this base width, should provide a reference for A-A collisions.

\item Conditional probabilities: Analysis of measure conservation in elementary collisions  \cite{epem} makes use of conventional conditional probabilities and generates interpretable results. The `conditional probability' invoked in \cite{bass} is not a probability. The binning geometry is not that of a conditional probability corresponding to Bayes' theorem, and the concept of normalization invoked is nonstandard. The properties of the balance function are consequently nonintuitive.

\item Cooling and diffusion time: According to arguments in \cite{bass} diffusion and cooling by isentropic expansion drive the charge-dependent correlation length in opposite directions. If the correlation length decreases it is therefore not clear what to conclude. Diffusion increases monotonically with the duration of hadronic rescattering. In \cite{bass} cooling by isentropic expansion occurs in the prehadronic phase. Cooling-induced narrowing would thus also signal a long-lived QGP. But this picture is at odds with a RHIC scenario of chemical decoupling at $\sim 170$ MeV followed by {\em hadronic} isentropic cooling and kinetic decoupling at $\sim 120$ MeV. Width reduction by hadronic cooling should signal an extended hadronic interval, not a long-lived QGP.

\item Balance function width and correlation length: The balance function was formulated by analogy with previous measures of correlation lengths in elementary collisions. It is thus implied that the balance function width is monotonically related to the charge-dependent correlation length on rapidity, if not equivalent to it. This implication is not valid.

\end{itemize}

\section{Charge correlations in elementary collisions}

A series of correlation measurements on rapidity in $e^+$-$e^-$, $p$-$p$ and $p$-$\bar p$ collisions was critical to formulating the phenomenological representation of nonperturbative QCD contained in the Lund model, and for verifying that quarks carry charge and flavor quantum numbers. The experimental techniques developed in those analyses were invoked to motivate the balance function. It is useful to examine them in detail with reference to applications in A-A collisions. 

\subsection{Techniques}

\begin{figure}[h]
\begin{tabular}{cc}
\begin{minipage}{.57\linewidth}
\epsfysize .49\textwidth
\xfig{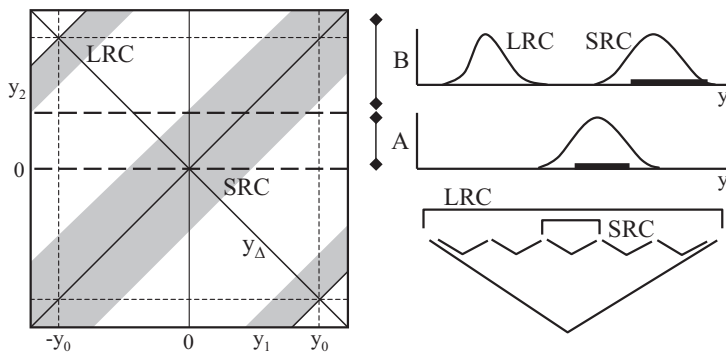}  
\end{minipage} &
\begin{minipage}{.37\linewidth}
\epsfysize 1\textwidth
\caption{Typical correlation analysis of elementary collisions. $y_0$ is a kinematic limit. Gray bands represent long-range and short-range measure conservation on rapidity difference $y_\Delta$. Projections on one marginal axis are shown for conditions (A,B) on the other, with a cartoon of the system space-time evolution.\label{src}}
\end{minipage}
\end{tabular}
\end{figure}

Correlations of conserved measures (charge, baryon number and strangeness) on azimuth and axial rapidity (or jet thrust axis in $e^+$-$e^-$ collisions) were studied extensively for elementary collisions \cite{epem}. Several closely-related quantities were defined to measure two-particle rapidity correlations: associated charge density (PEP4), rapidity differences with tagged intervals (OPAL) and charge compensation probabilities (TASSO), all based on the conditional-probability concept. A schematic of a typical analysis is shown in Fig. \ref{src}.
To accommodate limited statistics the two-particle or joint distribution on rapidity is typically divided into two regions by conditions placed on the rapidity of one (test or tagged) particle. The joint distribution for particles of opposite `charge' (Q, B, S) is then projected onto the rapidity for a second particle. These techniques can be compared with the structure implied by Bayes' theorem as illustrated in Fig. \ref{condition} (left panel). The conditional projections on rapidity typically contain peaked features reflecting correlation structure on rapidity difference. The locations, widths, amplitudes and polarities of the peaks reveal long-range and short-range measure conservation among Q, B and S on a `string' or 1D QCD color field.

\subsection{Results}

The correlation structure typically has two peaked components, a short-range component (SRC) due to local measure conservation during string fragmentation, and a long-range component (LRC) representing the conserved quantum numbers of the primary quark or nucleon pair, still localized near the initial or limiting rapidities ($\pm y_0$) in elementary collisions. The LRC reveals how the initial-state ($q$-$\bar q$, $p$-$\bar p$) quantum numbers (B, Q or S) propagate from the kinematic limit inward toward midrapidity (part of equilibration). The SRC reveals in part how intermediate-state quantum numbers created locally in pairs propagate outward from the formation region (also part of equilibration). 

Correlations in elementary collisions revealed that a 1D string has a natural correlation length for hadron pair correlations (SRC) $\sim$ 1 fm ({\em i.e.,} one hadron diameter) \cite{lund}. The fragmentation picture is consistent with hadrons initially forming a {\em contiguous} arrangement (a dense 1D fluid). Observed short-range charge (baryon number, strangeness) correlations reflect hadron `charge ordering' on a statistical basis: in the 1D system it is more probable to find alternating charge signs, a form of local measure conservation.

\subsection{Implications for the balance function and A-A collisions}

Comparison of p-p collisions with A-A centrality dependence is critical to many aspects of heavy ion collisions, including charge-dependent correlations. According to \cite{bass} `hadronization is nearly instantaneous' in p-p collisions. Whether this means the interval from projectile contact to hadronization, the hadronization process itself or the time from hadronization to kinetic decoupling is not clear. The last is nominally accessible to the balance function. 

Assuming the last interpretation the SRC rapidity correlation length in p-p collisions should be determined by the thermal velocity distribution, the Bjorken-expansion flow velocity and the characteristic length of 1D spatial hadron correlations at hadronization; the balance function width in p-p should acquire a minimum value which A-A collisions with `late' hadronization would approach asymptotically. In \cite{bass} is the statement `...compared to p-p collisions one expects the peak in the balance function in nucleus-nucleus collisions to be narrower....' which appears to be inconsistent. 

In \cite{bass} it is assumed that in A-A collisions the hadron pair-separation length scale at formation is zero, inconsistent with the fragmentation picture developed from elementary collisions. Any nonzero width of rapidity correlations should then be attributed only to diffusion and thermal velocities. Hadron diffusion in elementary collisions is said to be negligible, yet the charge-dependent correlation length is expected to be larger, again seemingly inconsistent. 

Systematics of hadron production in elementary collisions provide some initial guidance for A-A collisions, but the presence of a dense medium and the dimensionality difference of the hadronization process may have dramatic impact. The string system with its axial symmetry yields a compact formulation of 1D charge-correlation systematics. A similar description for hadronization in 2-3 dimensions in A-A would require a more general analysis of charge correlations in 3D momentum space. Restricting to a single momentum component in A-A analysis by analogy with elementary collisions could be misleading.

Invoking the balance function argument in its literal details we might predict an increase in balance-function width with increasing A-A centrality, from a p-p minimum to a maximum for mid-peripheral A-A due to increased diffusion time. An extended QGP duration in the more central events would then result in a reduction of the width toward the p-p limiting value.

\section{Balance Function Definitions}

The balance function is nominally defined as a linear combination of conditional probabilities: the probability that a particle of charge $a$ found within a condition interval on rapidity is accompanied by a particle of charge $b$ separated from $a$ by a specified rapidity difference. I present three equivalent definitions, the first as presented in \cite{bass,jeon}, the second based on binning the single-particle rapidity distribution, the third in terms of properly-defined conditional probabilities. These definitions form the basis for algebraic study of the balance function.

\subsection{Theory definition}

This definition was formulated by analogy with analysis of elementary collisions in terms of joint and marginal rapidity distributions. Using the notation in \cite{jeon}, $\Delta_1,\Delta_2$ in Eq.~(\ref{jeon}) are defined as conditions on some combination of the rapidity variables of the joint distribution
\bea \label{jeon}
B(\Delta_1 : \Delta_2 ) &\equiv& {1 \over 2} \, \left\{ {N_{-+}(\Delta_1 , \Delta_2) - N_{++}(\Delta_1, \Delta_2) \over N_+(\Delta_2)} +  {N_{+-}(\Delta_1 ,\Delta_2) - N_{--}(\Delta_1 ,\Delta_2) \over N_-(\Delta_2)} \right\} \\ \nonumber
&=& {1 \over 2} \, \left\{ \sum_{a,b=-}^+ -ab {N(ab,\Delta_1 , \Delta_2) \over N_b(\Delta_2)} \right\}
\eea
where $\Delta_1 : \Delta_2$ is read pairs with condition $\Delta_1$ given that condition $\Delta_2$ is satisfied. Each $N_b(\Delta_2)$ or $N_{ab}(\Delta_1,\Delta_2)$ is an ensemble-averaged event-wise number of particles of charge $b$ or pairs of charge combination $ab$ respectively, satisfying the rapidity combination $(\Delta_1, \Delta_2)$. The conditions are taken to be a rapidity acceptance and a rapidity difference respectively: $\Delta_2 \rightarrow \Delta y, ~ \Delta_1 \rightarrow k \delta y$. In particular, domains $\Delta_1$ are diagonal strip integrals on the joint rapidity distribution that differ markedly from the usual conditional probability definition.

\subsection{Definition based on uniform binning}

A binned rapidity axis $y$, with acceptance $\Delta y$, bin size $\delta y$ and bin number $M(\Delta y,\delta y)$, supports a particle distribution. Bin $i$ has integrated multiplicity $n_{ai}(\delta y)$ of `charge' species $a$. The total multiplicity of charge $a$ in the acceptance is $\sum_{i=1}^{M(\Delta y,\delta y)} n_{ai}(\delta y) \equiv N_a(\Delta y) $. The bin counts represent integrals of the form
\bea
\bar n_{ai}(\delta y) \approx \int_{y_i - \delta y / 2}^{y_i + \delta y / 2}  \rho_{1n_a}(y) \, dy
\eea
where number density $\rho_{1n_a}(y)$ represents a single-particle inclusive parent distribution estimated by the histogram of ensemble averages $\{\bar n_{ai}(\delta y)\}$ and from which each event provides a sample $\{ n_{ai}(\delta y)\}$. A similar description holds for pair counts from the two-particle distributions. The balance function histogram $\{B_k\}$ is then defined as
\bea \label{baldef}
B_k(\Delta y,\delta y) &\equiv& 1/2 \, \sum_{a,b=-}^+  -ab \,{ 1 \over \sum_{i=1}^M \, \bar n_{ai}} \,\sum_{{i=1}}^{M(\Delta y,\delta y)-k} \,  \overline{ n_{ai} \cdot (n_{b(i+k)} - \delta_{ab}\delta_{k0})  }   \hspace{.1in} k \in [0,M-1] \\ \nonumber
& \equiv& 1/2 \, \sum_{a,b=-}^+  -ab \, {1  \over \sum_{i=1}^M \, \bar n_{ai} }\sum_{{i=1-k}}^{M(\Delta y,\delta y)} \,   \overline{ n_{ai} \cdot (n_{b(i+k)} - \delta_{ab}\delta_{k0})  }   \hspace{.3in} k \in [1-M,0]
\eea
where the delta functions represent omission of self-pairs.
This definition is equivalent to those in \cite{bass,jeon} but makes explicit the geometry of the strip integrals, and facilitates subsequent comparison with the autocorrelation distribution.

\subsection{Definition based on conditional probabilities}

\begin{figure}[h]
\begin{tabular}{cc}
\begin{minipage}{.64\linewidth}
\epsfysize .44\textwidth
\xfig{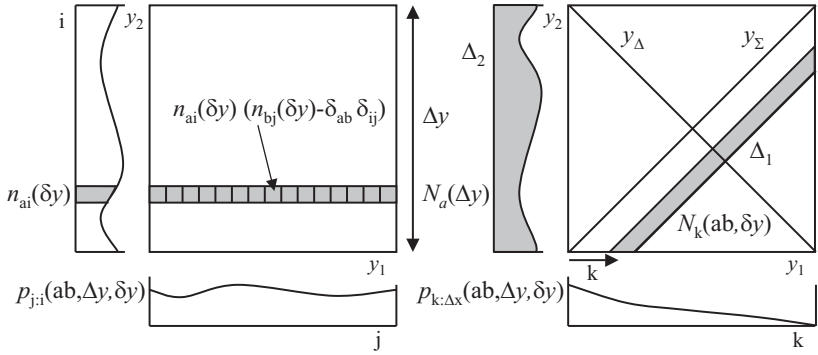}  
\end{minipage} &
\begin{minipage}{.3\linewidth}
\epsfysize 1\textwidth
\caption{Standard conditional probability definition (left panel) according to Bayes' theorem and definition associated with balance function (right panel).\label{condition}}
\end{minipage}
\end{tabular}
\end{figure}

Basic two-point and single-point probabilities can be defined in terms of joint and marginal bin multiplicities as $p_{ij}(ab) = \overline{n_{ai}(n_{bj} - \delta_{ab}\delta_{ij})} / \overline{ N_a( N_b-\delta_{ab})}$ and $p_i(a) = \overline{ n_{ai}( N_b-\delta_{ab}) }/ \overline{ N_a( N_b-\delta_{ab})}$. Bayes' theorem for ensemble means is then
\bea
p_{j:i}(ab,\Delta y,\delta y) \equiv {p_{ij}(ab) \over p_i(a)} = {\overline{n_{ai}(n_{bj} - \delta_{ab}\delta_{ij})} \over \overline{ n_{ai}( N_b-\delta_{ab}) }}
\eea
This is the conditional probability that a particle of charge $b$ will appear in bin $j$ given that bin $i$ is occupied by a particle of charge $a$ as determined by the joint distribution $ \overline{n_{ai}(n_{bj} - \delta_{ab}\delta_{ij})}$, as shown in the left panel of Fig. \ref{condition}, of which distribution $\overline{ n_{ai}( N_b-\delta_{ab}) }$ is a marginal projection. Elementary collisions analyses adopted this conventional conditional-probability technique, projecting onto one of the marginal axes with a condition imposed on the other (typically dividing the accessible rapidity into two bins) as shown in Fig. \ref{src}. The result provides simple interpretation of projected peak widths in terms of correlation lengths. The cost of the method is a skewed projection of the underlying autocorrelation distribution onto the marginal axis, a nonoptimal compression technique. The technique is however commensurate with the low statistics available in early analyses. For A-A collisions a more precise treatment is justified.

The conditional geometry of the balance function is modified as shown in Fig. \ref{condition} (right panel), with the equivalences $n_{ai}(\delta y) \rightarrow N_a(\Delta y)$, $n_{ai}(n_{bj} - \delta_{ab} \delta_{ij}) \rightarrow N_k(ab,\delta y) \leftrightarrow N_{ab}(\Delta_1,\Delta_2)$. $\bar N_k(ab) \equiv \sum_i \overline{n_{ai}\, (n_{b(i+k)}-\delta_{ab}\delta_{k0})}$ is a strip integral, a projection onto the difference variable $y_\Delta$. The conditional probability according to the balance function treatment is then
\bea
  p_{k:\Delta y}(ab)  \equiv \bar N_k(ab,\Delta y,\delta y) / \overline{  N_a(\Delta y) \, (N_b(\Delta y) - \delta_{ab}) }
\eea
With this definition the balance function takes the form
\bea \label{balcond}
B_k(\Delta y,\delta y) &\equiv& {1 \over 2} \left\{ \sum_{a,b=-}^+ -ab \, p_{k:\Delta y}(ab) \, \overline{  N_a(\Delta y) \, (N_b(\Delta y) - \delta_{ab}) } /  \bar N_a(\Delta y) \right\} \\ \nonumber
&\equiv&  {1 \over 2} \left\{ \sum_{a}  p_{k:\Delta y}(aa) -  \sum_{a,b} ab \,  p_{k:\Delta y}(ab) \,\overline{  N_a(\Delta y) \, N_b(\Delta y) } /  \bar N_a(\Delta y) \right\}
\eea
with $\sum_k p_{k:\Delta y}(ab,\Delta y,\delta y) \equiv 1$ for each $ab$ combination by construction.

\section{The Autocorrelation and the Balance Function}

The autocorrelation, a special case of the cross correlation, is the projection of a two-point distribution onto its difference variables, a good compression strategy if correlations are approximately space invariant (invariant on sum variable $y_\Sigma$ in Fig. \ref{condition}). It is a well-established correlation measure used in time-series analysis for many decades \cite{auto1,auto2}. The autocorrelation characterizes the internal correlations of a single distribution. The term `cross-correlation' applies to mixed distributions, for instance of different charge types. For simplicity I refer to cross- and autocorrelations both as autocorrelations. A {\em correlation length} denoted by $\xi_y$ typically refers to a characteristic length or width of the autocorrelation distribution (on the difference variable $y_\Delta$ in Fig. \ref{condition}).

\subsection{Autocorrelation definition}

In terms of 2D bin integrals of the joint distribution the autocorrelation for an {\em aperiodic} distribution on a bounded domain \cite{auto2} (also called an {\em unbiased sample} autocorrelation) applicable to {\em e.g.,} distributions on $y, \,  \eta$ is
\bea \label{autodef}
A_{k,obj}(ab) &\equiv& {1  \over M-k } \sum_{{i=1}}^{M(\Delta y,\delta y)-k} \,   \overline{ {n_{ai} \cdot (n_{b(i+k)} - \delta_{ab}\delta_{k0}) } } ~~~~k \in [0,M-1] \\ \nonumber
& \equiv& {1  \over M+k } \sum_{{i=1-k}}^{M(\Delta y,\delta y)} \,   \overline{ {n_{ai} \cdot (n_{b(i+k)} - \delta_{ab}\delta_{k0}) } }   ~~~~~~~k \in [1-M,0]
\eea
For simplicity in what follows I suppress expressions for negative values of $k$. The autocorrelation for {\em periodic$\,$}  distributions on $\phi$ is $A_{k,obj}(ab) \equiv {1  / M}\, \sum_{{i=1}}^{M} \,   \overline{ {n_{ai} \cdot (n_{b(i+k)} - \delta_{ab}\delta_{k0}) } } $, and is symmetric about the midpoints $|k| = (M - 1)/2$ for real $n_i$. The autocorrelation (omitting self pairs) is thus composed of factorial moments and cross moments at the binning scale $\delta y$, with the weighted integral
\bea
\sum_{k=1-M}^{M-1} (M - k) A_{k,obj}(ab) &=& \overline{ N_a(\Delta y)\,  (N_b(\Delta y) - \delta_{ab})}
\eea
It follows that
\bea \label{form}
\overline{N_a(\Delta y)\,  (N_b(\Delta y) - \delta_{ab})}\, p_{k:\Delta y}(ab) \equiv (M - k) A_{k,obj}(ab)
\eea

\subsection{Autocorrelation reference and net autocorrelation}

The differential or {\em net} autocorrelation depends on a reference definition. For this study I assume a factorization reference constructed from the inclusive single-particle distribution
\bea
A_{k,ref}(ab) &\equiv& {1 \over M - k} \, \sum_{i=1}^{M-k} \, \bar n_{ai}  \, \bar n_{b(i+k)} 
\eea
with integral $\sum_{k=1-M}^{M-1} (M - k) A_{k,ref}(ab) = \bar N_a(\Delta y)\,  \bar N_b(\Delta y) $
and charge-independent (CI) and charge-dependent (CD) linear combinations
$  A_{k,ref}(CI) \equiv  \sum_{ab} \, A_{k,ref}(ab) 
= {1 \over M - k} \, \sum_{i=1}^{M-k} \, \bar n_{i}  \, \bar n_{i+k} 
\approx \bar n^2 $ and
$  A_{k,ref}(CD) \equiv  \sum_{ab} \, ab\,  A_{k,ref}(ab) 
= {1 \over M - k} \, \sum_{i=1}^{M-k} \, \bar \epsilon_{Qi}\bar n_{i}  \, \bar \epsilon_{Q(i+k)} \bar n_{i+k}  \approx  0 $,  where bin-wise charge asymmetry $\bar \epsilon_{Qi} = \bar Q_i / \bar n_{i} \equiv  \sum_{a=-}^+ a \bar n_{ai} / \sum_{a=-}^+ \bar n_{ai}$.

The net autocorrelation is defined as $ \Delta A_k(ab) \equiv  A_{k,obj}(ab) - A_{k,ref}(ab) $ with CI and CD linear combinations. I decompose the CD net autocorrelation into its integral (the net-charge variance at scale $\delta y$), and the zero-integral CD net autocorrelation $\Delta \hat A_k(CD)$
\bea
\Delta A_k(ab) &\equiv& \Delta \hat A_k(ab) + {{\overline{ N_a \,( N_b- \delta_{ab})} - \bar N_a \, \bar N_b } \over \bar N_a \,  \bar N_b } A_{k,ref}(ab) \\ \nonumber
\Delta A_k(CD) &\simeq& \Delta \hat A_k(CD) + \left[{\sigma^2_Q(\Delta y) \over \bar N(\Delta y) } - 1\right] \cdot{A_{k,ref}(CI) \over \bar N(\Delta y) } \\ \nonumber
&\approx &\Delta \hat A_k(CD) + \sigma^2_Q(\delta y) - \bar n / M
\eea
The zero-integral CD net autocorrelation $\Delta \hat A_k(CD)$ records charge-dependent {\em pair transport} within events, averaged over the ensemble. This decomposition clarifies the role of net-charge fluctuations in the CD net autocorrelation and the balance function. This particular form of net-charge variance results from the assumption of a factorization reference, in which case the variance agrees with the elementary statistical definition. This is usually not the best reference choice for a precision analysis.

\subsection{Relating the balance function to the autocorrelation}

The balance function is a hybrid between a conditional probability and an autocorrelation. Its definition uses the language of conditional probabilities but incorporates a form of projection onto a difference variable in the manner of an autocorrelation. The two distributions can be related algebraically according to the above definitions (Eqs.~(\ref{baldef}) and (\ref{autodef}) for balance function and autocorrelation in terms of binning and  Eq.~(\ref{form}) for the form factor)
\bea \label{balauto}
B_k(\Delta y,\delta y) &\equiv&  1/2 \, \sum_{a,b=-}^+  -ab \, { (M - k)\, A_{k,obj}(ab) \over  \bar N_{a} } \\ \nonumber
&=& {(M - k) \, A_{k,ref}(CI) \over \bar N^2 } \, \left\{- {\bar N \,A_{k,ref}(CD) \over A_{k,ref}(CI) } -{\bar N \, \Delta A_{k}(CD)   \over A_{k,ref}(CI) }   \right\}\ \\ \nonumber
&\simeq&  p_{k:\Delta y}(CI) \, \left\{ 1 - {\sigma^2_Q(\Delta y) \over \bar N_{}(\Delta y)} -  {\bar N \,  \Delta \hat A_{k}(CD) \over A_{k,ref}(CI) } \right\} \\ \nonumber
\eea
where $\Delta \hat A_{k}(CD) / A_{k,ref}(CI) \ll 1$ is a correlation measure per final-state hadron {\em pair}. The additional factor $\bar N$ gives the correlation measure per charged hadron. The overall sign convention for the balance function is opposite to isospin convention. 

The unit-integral first factor is approximated by the triangle $(1 -  |k|/M) / M$ if the single-particle $dN/d\eta$ distribution is nearly uniform within the acceptance.  In \cite{bass} a width reference is suggested: `In a globally equilibrated system...there would exist no correlation between the balancing charges, and the...width of the balance function would then correspond to the extent of single-particle emission in momentum space.' We find that the reference distribution is actually a normalized projection of the $d^2N/dy_1 dy_2$ pair distribution (as seen through the detector acceptance) onto the difference variable $y_\Delta \approx k \delta y$. The triangular shape of this projection is a source of difficulty in interpreting the balance function. The first two terms in the second factor represent net-charge fluctuations (integral of the CD net autocorrelation) in the acceptance, with canonical suppression. The zero-integral last term represents CD charge-pair transport, which may be characterized by a correlation length if the net autocorrelation is peaked near $k = 0$.

\subsection{Balance function integral}

From the definition in terms of conditional probabilities in Eq.~(\ref{balcond}) we obtain
\bea \label{balint}
\sum_k B_k(\Delta y,\delta y) &=&   1 -  {1 \over 2}\sum_{a,b=-}^+ ab {\overline{N_a(\Delta y) N_b(\Delta y) } \over \bar N_a(\Delta y)} \\ \nonumber
&\approx& 1 - {\sigma^2_Q(\Delta y) \over \bar N_{}(\Delta y)}
\eea
which also follows directly from Eq.~(\ref{balauto}). The accuracy of the approximation depends on charge asymmetry \mbox{$\bar \epsilon_Q \ll 1$}, 
a condition well satisfied for  QED charge and central A-A collisions near midrapidity. For baryon number, strangeness or noncentral collisions this condition may not be satisfied, as noted in \cite{jeon}.
\begin{figure}[h]
\begin{tabular}{cc}
\begin{minipage}{.47\linewidth}
\epsfysize .5\textwidth
\xfig{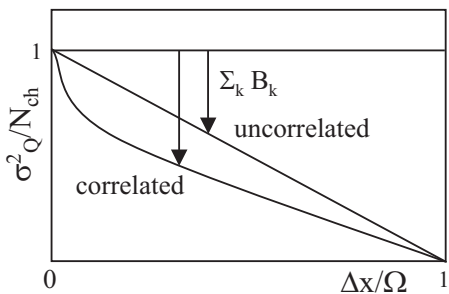}  
\end{minipage} &
\begin{minipage}{.47\linewidth}
\epsfysize 1\textwidth
\caption{Relationship between balance-function integral and net-charge fluctuations plotted $vs$ acceptance fraction of $\Omega \equiv 4\pi$. Sloping line shows the effect of canonical suppression without charge correlations. Curve suggests the effect of additional charge correlations.\label{sigq}}
\end{minipage}
\end{tabular}
\end{figure}
The balance-function integral thus depends on the acceptance fraction of the total $4\pi$ charge distribution in the form of canonical suppression of net-charge fluctuations and nontrivial charge correlations, as shown in Fig. \ref{sigq}. For a $4\pi$ detector $\sigma^2_Q \equiv 0$ and $\sum_k B_k \equiv 1$.

The appearance of an extraneous variance in a correlation measure is an example of measure bias, a cross coupling of one aspect of correlation analysis with another, producing a measure that is difficult to interpret. Equivalently, the balance function in Eq. (\ref{balauto}) can be seen as an {\em integral equation} for the charge-dependent correlations represented by the zero-integral CD net autocorrelation. Charge correlations enter twice: once in the integral $\sigma^2_Q / \bar N$ and once in the detailed shape of the pair-transport distribution $\Delta \hat A_k$.  The latter is by itself an example of an unbiased measure which reveals the fundamental correlation length and amplitude in an unambiguous manner.

\subsection{Balance function reference}

The balance function distribution should be compared to a reference representing both absence of charge-dependent correlations and conservation of net charge within the detector acceptance. The factorization reference used for this autocorrelation (and hence balance function) illustration was adopted to simplify illustration. An alternative reference can be generated directly from data by a mixing procedure.

Event mixing in this case involves charge mixing and/or rapidity mixing. Charge mixing does not respect conservation of net charge in $4\pi$, which is a constraint for real data. The net-charge variance then takes the grand-canonical limit $\sigma^2_Q / \bar N \approx 1$, driving the integral of the balance function to zero as seen in Eq.~(\ref{balint}) and Fig. \ref{sigq}. Forming mixed events by selecting random particles from an event ensemble also mixes charges and produces a reference that doesn't respect conservation of net charge. Rapidity mixing randomizes the {\em location} of existing charges within the acceptance, which reduces $\Delta \hat A_k(CD)$ to zero in Eq.~(\ref{balauto}). The resulting reference retains the correct form factor and net-charge fluctuation term. The algebraic form representing rapidity mixing is
\bea
B_{k,ref}(\Delta y,\delta y) &\simeq&  p_{k:\Delta y}(CI) \, \left\{ 1 - {\sigma^2_Q(\delta y) \over \bar N_{}(\delta y)} 
\right\} \\ \nonumber
\eea
which can be compared to a balance function derived from unmixed real data to reveal the nontrivial charge correlations represented by ${\bar N \,  \Delta \hat A_{k}(CD) / A_{k,ref}(CI) }$.

\subsection{Pair-ratio distributions}

One can also generate ratios of object and reference pair distributions for each charge combination to access the CD net autocorrelation directly \cite{aya}. Event-wise ratios of sibling-pair bin counts to mixed-pair bin counts are constructed, forming sibling pairs within single events and mixed pairs by matching each object event to its nearest neighbor in a space spanned by global event properties such as total multiplicity \cite{plaid} and then averaging the event-wise bin ratios over an event class
\bea
\bar r_{ij}(ab) =   \overline{ \{n_{ai}\, (n_{bj} - \delta_{ab}\delta_{ij})\}_{sib}\, / \,  \{ n_{ai}\, (n_{bj} - \delta_{ab}\delta_{ij}) \}}_{mix}
\eea
The relative net autocorrelation is expressed in terms of this ratio distribution by
\bea
\Delta \hat A_{k}(ab) / A_{k,ref}(ab) &\equiv&   \left[ {1  \over M-k } { \sum_{{i=1}}^{M(\Delta y,\delta y)-k} \, \bar r_{i,i+k}(ab)   } \right] -1 
\eea
Because the sibling and mixed pair totals are matched on an event-wise basis the net autocorrelation $\Delta \hat A_k(ab)$ has zero integral by construction for each $ab$ combination. This method accesses the 2D correlation structure of {\em each} charge combination, not only the CD or CI combination. The CD linear combination reveals the charge-dependent correlations which are the nominal object of the balance function. The pair-ratio method is acceptance independent (whereas $\sigma^2_Q$ and B are not) and provides direct access to the basic correlation length $\xi_y$ and amplitude $a_0$, as well as other details of CD pair transport.

\section{Balance Function Width Systematics}

The theoretical hypothesis which motivates the balance function is expressed in terms of parametric variation of a charge-dependent correlation length. The balance function is said to represent this correlation length through its width. I examine the relationship between balance function width and correlation length with a simple model that corresponds closely to correlation structure observed in A-A collisions.

\subsection{Correlation model}

The model is formulated in terms of ratio $\Delta \hat A_k(CD) / A_{k,ref}(CI)$ derived from a nominally uniform distribution of charges on momentum variable $y$ (which could be rapidity $y$, pseudorapidity $\eta$ or azimuth $\phi$). Given difference variable $y_\Delta \equiv y_1 - y_2$ I define normalized difference variable $x \equiv y_{\Delta} / \Delta y \in [-1,1]$ and normalized correlation length $x_0 \equiv \xi_{y} / \Delta y \in [0,\infty)$, both relative to detector acceptance $\Delta y$. This continuum model relates to the discrete case by $k/M \rightarrow k \delta y / M \delta y \rightarrow x$. I adopt a simple exponential form for the relative net autocorrelation consistent with observations: $- \Delta \hat A(x) / A(x) = a_0 \, \{ e^{-|x|/x_0} - \Delta_0\}$. The integral condition
\bea
\int^{1}_{-1}  (1 - |x|) \, \Delta \hat A(x) / A(x) \,dx = 0
\eea
 is imposed to determine the offset $\Delta_0$, from which we obtain
\bea \label{off}
\Delta_0 = 2x_0 \, \{ 1 - x_0 \, (1-e^{-1/x_0}) \}
\eea

\begin{figure}[h]
\begin{tabular}{cc}
\begin{minipage}{.57\linewidth}
\epsfysize .5\textwidth
\xfig{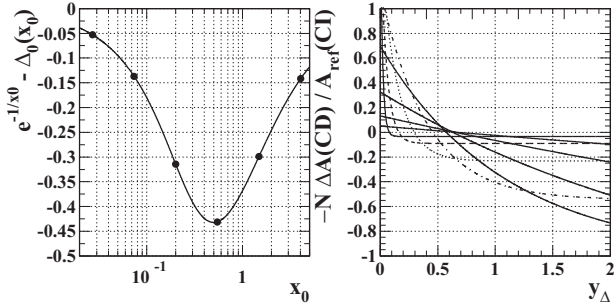}  
\end{minipage} &
\begin{minipage}{.37\linewidth}
\epsfysize 1\textwidth
\caption{Autocorrelation distributions (right panel) for a selection of eight correlation lengths with amplitude held fixed \cite{aya} and corresponding distribution values (left panel) at $x = 1$ (two correlation lengths are outside the plot range). \label{autocorr}}
\end{minipage}
\end{tabular}
\end{figure}
The net-autocorrelation model is shown in Fig. \ref{autocorr} (right panel) for a selection of eight relative correlation lengths $x_0$ plotted $vs$ rapidity difference assuming $\Delta y = 2$, with amplitude $a_0$ held fixed and offset $\Delta_0$ varying according to Eq.~(\ref{off}) to maintain the zero-integral condition. Distribution values at $y_\Delta = 2$ or $x = 1$ are shown in the left panel to illustrate the systematic variation of $\Delta_0$. Defining $A_0 \equiv \bar N \, a_0$ and $Q_0 \equiv \sigma^2_Q / \bar N$,
the balance function for this idealization is then 
\bea
B(\delta y,x) / \delta y &\equiv& {(1-|x|)}  \left\{1 - {\sigma^2_Q \over \bar N} - {\bar N\, \Delta \hat A(CD;x) \over A_{ref}(CI;x)} \right\} \\ \nonumber
&=& {(1-|x|)}  \left\{1 - Q_0 + A_0 \, \left( e^{- |x| / x_0} - \Delta_0   \right)  \right\}  
\eea
with the condition $\int_{-1}^1 B(\delta y,x) \,dx = 1-Q_0(\Delta y)$. 
The basic correlation parameters $a_0$ and $\xi_y$ are best obtained directly from the CD net autocorrelation, from which we can derive other related quantities, in particular the balance function.

\subsection{Width systematics}

\begin{figure}[h]
\begin{tabular}{cc}
\begin{minipage}{.57\linewidth}
\epsfysize .5\textwidth
\xfig{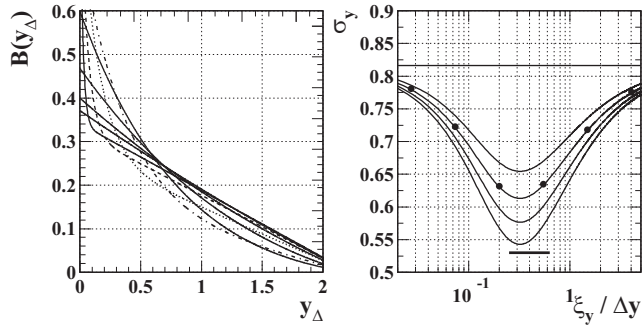}  
\end{minipage} &
\begin{minipage}{.37\linewidth}
\epsfysize 1\textwidth
\caption{Balance functions (left panel) corresponding to eight autocorrelations in Fig. \ref{autocorr} and distributions of $rms$ widths (right panel) for four amplitudes ($A_0$) in Eq.~(\ref{param}).  The dots correspond to balance functions in the left panel (two outside the plot range). \label{balcorr}}
\end{minipage}
\end{tabular}
\end{figure}

We now investigate the parametric dependence of the balance function width on the model parameters -- the amplitude and correlation length. I define the mean-square width of the balance function as
\bea \label{param}
\sigma^2_x &=& {1 \over 1 - Q_0} \int_{-1}^1 \, x^2 \, B(x) \, dx  \\ \nonumber
&=&  {1 \over 1 - Q_0}\int_{-1}^1 \, x^2 \, (1-|x|)  \,   \left\{1 - Q_0 + A_0 \, \left( e^{- |x| / x_0} - \Delta_0   \right)  \right\} \, dx  \\ \nonumber
&=& {1 \over 6} + {2A_0\over 1 - Q_0} \left[ (2x_0^3 - 6x_0^4 )\, (1-e^{-1/x_0}) + (6x_0^3 + x_0^2) \, e^{-1/x_0} - {1 \over 12} \, \Delta_0 \right] 
\eea
with $\sigma_y = \Delta y \cdot \sigma_x$ being the corresponding $rms$ width on rapidity difference. The $rms$ width for the case of no correlations ($A_0 = 0$) is $\sqrt{1/6} \approx 0.4$ times $\Delta y$. For the STAR detector the latter is about 2.1, and the uncorrelated or rapidity-mixed reference $rms$ is observed to be about 0.83 as shown in Fig. \ref{balcorr} (right panel).

The balance functions in the left panel exhibit the expected behavior: charge correlation structure deviating from a triangular reference distribution whose area is determined by net-charge fluctuations within the acceptance. The surprising result is in the right panel. The horizontal line at 0.83 approximates the reference $rms$ width corresponding to no charge correlations. The balance function $rms$ width otherwise has a minimum for $\xi_y / \Delta y \approx 1/3$, which coincides with typical values for heavy ion collisions and the STAR detector (lower horizontal bar) \cite{aya}. For relative correlation lengths much below this value the width of the balance function would actually anticorrelate with the charge-dependent correlation length. For heavy ion collisions the balance function $rms$ width is thus dominated by the correlation {\em amplitude}, as illustrated by the four curves in the right panel. It is essentially insensitive to the charge-dependent correlation length.

\section{Conclusions}

The balance function, proposed as a method to probe QGP existence and duration by means of charge-dependent correlations, has stimulated re-examination of local measure conservation in heavy ion collisions and its analysis methods. I present here several critical issues concerning the motivating chain of argument and interpretation of the balance function as a correlation measure.

If the charge-dependent net autocorrelation has a simple peaked structure near zero momentum difference (`short-range' correlations, as observed in both elementary and A-A collisions) the basic charge-dependent correlation parameters are the peak amplitude and its $rms$ width, the latter conventionally called the correlation length. Detailed aspects of the distribution are also important, but for the balance function hypothesis these are the relevant quantities. The purpose of a correlation analysis applicable to the physics of the balance function should be inference of this correlation length and amplitude.

The balance function apparently confuses amplitude and correlation length in its width definition. For heavy ion collisions viewed by a large-acceptance detector (STAR at RHIC) the balance function width is ironically insensitive to the correlation length, is dominated instead by the correlation amplitude.  Theoretical hypotheses addressing the charge-dependent correlation length in terms of  the balance function width are consequently misdirected. 

The supporting theoretical chain of argument concerning initial charge correlations at hadronization, exclusive influence of diffusion and thermal velocity on final-state correlation lengths and relation of A-A correlations to those in elementary collisions seems iconsistent. The structure of hadronization in A-A collisions is an open question; extrapolations from 1D string phenomenology are unwarranted without careful re-examination. The geometry of the prehadronic medium at hadronization may dominate the structure of A-A charge correlations.


\section{Acknowledgements}

I appreciate helpful discussions with R.L. Ray (U. Texas, Austin). This work was supported in part by USDOE contract  DE-FG03-97ER41020.

\end{document}